\documentclass[journal]{IEEEtran}
\usepackage{etex}
\reserveinserts{100}
\usepackage{mathtools}
\usepackage[dvipsnames]{xcolor}
\usepackage{graphics, theorem, times, amsfonts, graphicx, amssymb, cite}
\usepackage[outdir=./]{epstopdf}
\usepackage{tikz}
\usetikzlibrary{shapes,arrows}
\usetikzlibrary{matrix} 
\usetikzlibrary{decorations.markings} 
\usetikzlibrary{calc} 
\usetikzlibrary{decorations.pathmorphing,snakes, positioning, decorations.pathreplacing}

\usetikzlibrary{fit}
\usepackage{pgfplots}
\usepackage{color}
\usepackage{setspace}
\usepackage{hyperref}
\usepackage{multirow}
\usepackage{rotating}
\usepackage{comment}
\usepackage{caption}
\usepackage{subcaption}
\usepackage[affil-it]{authblk}
\usepackage{bm}
\usepackage{algorithm}
\usepackage{algorithmic}
\usepackage{enumerate}
\usepackage{morefloats}
\usepackage{setspace}

\input{mysymbol.sty}
\usepackage{needspace}

\newcommand{\QED}{\hfill\ensuremath{\blacksquare}}

\def\E{\mathbb{E}}
\def\P{\mathbb{P}}

\def\Tr{\text{Tr}}

\newtheorem{proposition}{\hspace{0pt}\bf Proposition}

\newtheorem{remark}{\hspace{0pt}\bf Remark}

\title{Control Aware Radio Resource Allocation in Low Latency Wireless Control Systems}
\author{Mark Eisen$^*$ \quad Mohammad M. Rashid$^\dagger$ \quad Konstantinos Gatsis$^*$ \\ \textup{Dave Cavalcanti$^\dagger$\quad Nageen Himayat$^{\dagger}$ \quad Alejandro Ribeiro$^*$}
\thanks{Supported by the Intel Science and Technology Center for Wireless Autonomous Systems and ARL DCIST CRA W911NF-17-2-0181. The authors are with the ($^*$)Department of Electrical and Systems Engineering, University of Pennsylvania and ($^\dagger$)Wireless Communications Research, Intel Corporation. Email: maeisen@seas.upenn.edu, mamun.rashid@intel.com, kgatsis@seas.upenn.edu, dave.cavalcanti@intel.com, nageen.himayat@intel.com, aribeiro@seas.upenn.edu.}}

\begin{document}

\thispagestyle{empty}
\maketitle

\begin{abstract}
We consider the problem of allocating radio resources over wireless communication links to control a series of independent wireless control systems. Low-latency transmissions are necessary in enabling time-sensitive control systems to operate over wireless links with high reliability. Achieving fast data rates over wireless links thus comes at the cost of reliability in the form of high packet error rates compared to wired links due to channel noise and interference.  However, the effect of the communication link errors on the control system performance depends dynamically on the control system state. We propose a novel control-communication co-design approach to the low-latency resource allocation problem. We incorporate control and channel state information to make scheduling decisions over time on frequency, bandwidth and data rates across the next-generation Wi-Fi based wireless communication links that close the control loops.  Control systems that are closer to instability or further from a desired range in a given control cycle are given higher packet delivery rate targets to meet. Rather than a simple priority ranking, we derive precise packet error rate targets for each system needed to satisfy stability targets and make scheduling decisions to meet such targets while reducing total transmission time. The resulting Control-Aware Low Latency Scheduling (CALLS) method is tested in numerous simulation experiments that demonstrate its effectiveness in meeting control-based goals under tight latency constraints relative to control-agnostic scheduling.

\end{abstract}


\section{Introduction}\label{sec_intro}
The reliability and robustness of wireless communication systems has become a key component in the design and implementation of large scale systems in the Internet-of-Things (IoT). For these systems to operate successfully, it is important that the data collected by sensors can be communicated throughout the IoT network. While the most reliable form of data communication is through wired connections, the scale and mobility of modern IoT settings has rendered the cost of installing and maintaining a wired network a significant challenge~\cite{wollschlaeger2017future}. Consequently, there is great interest in the design of wireless control systems that can achieve reliable performance~\cite{varghese2014wireless, li2017review}. Low-latency wireless transmissions are however a necessary feature in many wireless IoT systems, particularly those in industrial control~\cite{varghese2014wireless}.  The primary challenge is then the inherent trade-off that occurs between reliability and latency, as it is difficult to maintain high reliability while using higher data rates due to the stochasticity of the wireless channel. It is necessary to design resource allocation and scheduling protocols for wireless control systems that can both meet reliability \emph{and} latency requirements of the control system.

In the wireless communications research and industry, many radio resource allocation schemes in the form of wireless scheduling techniques have been proposed to provide reliability, or quality of service (QoS), to users across the network in the form of throughput, fairness and/or latency~\cite{cao2001scheduling,wongthavarawat2003packet,fattah2002overview,yaacoub2012survey}. For instance, round-robin scheduling is a common approach due to its simplicity inherent fairness. Algorithms such as proportional fair (PF)~\cite{kim2005proportional}, max-SNR~\cite{knopp1995information} and their variations have been designed for maximizing the overall system throughput and/or fairness. For time-sensitive applications such as industrial control and automation, the experiences of individual users/devices are highly dependent on the ability of the network to deliver their packets within a latency bound. Generic delay-aware schedulers such as EDF~\cite{wu2014analysis} and WFQ~\cite{lu1999fair} do not adapt to wireless channel conditions, while channel state information is considered in M-LWDF \cite{andrews2001providing}. However, all such methods are unable to leverage the information on the control system dynamics and thus can make unsuitable radio resource allocation decisions when deployed in time-sensitive wireless control systems.

In the context of wireless control systems, there have been a range of works that incorporates control system information in the networking and communication policies.
The mechanisms usually examined are either static or dynamic. Typical examples of the former type are periodically protocols where the wireless devices transmit in a predefined repeating order, e.g., round-robin. Control system stability under such protocols can be analyzed -- see, e.g.,~\cite{Hespanha_survey, Schenato_foundations, Donkers_switched, Branicky_stability}. Periodic sequences leading to stability~\cite{Hristu_shared_feedback}, controllability and observability~\cite{Hristu_communication_control}, or optimizing control objectives~\cite{LeNy_resource_LQR,Meier_measurement_control, Scheduling_control_combinatorics} have been proposed.
Dynamic schedulers do not rely on a predefined sequence but decide access to the communication medium dynamically at each step. 
Initial approaches abstract control performance requirements in the time/frequency domain, e.g., how often a task needs resource access, employing algorithms from real-time scheduling theory~\cite{Branicky_RM, Liu_Real_time_systems}. More recent scheduling approaches often depend on the current control system states, i.e., informally the subsystem with the largest state discrepancy is scheduled to communicate -- see, e.g.,~\cite{Donkers_switched, Walsh_stability, Hristu_Kumar_interrupt_based, Egerstedt_queue, Hirche_Scheduling_Price, Cervin_event_scheduling, mamduhi2014event,shi2011optimal,han2017optimal}.
Alternatively scheduling can take into account current wireless channel conditions opportunistically to meet target control system reliability requirements~\cite{GatsisEtal15}. None of these approaches, however, are explicitly designed to achieve strong performance it low-latency scenarios.

In this paper, we develop a control-aware, high reliability, low-latency IEEE 802.11ax WiFi protocol~\cite{bellalta2016ieee} that is designed to reduce the total transmission time at each uplink cycle in the control loop. This is done through the mathematical formulation of the control system design goal in the form of a Lyapunov function that ensures stability of the control system. This formulation naturally induces a bound on the packet delivery rate each control system needs to achieve to meet the control-based goal. Such packet delivery rates depend upon current control and channel states and thus dynamically change over the course of the system life time. This can be viewed as an opportunistic protocol with respect to both the current control states \emph{and} channel conditions. Furthermore, because these control-based success rate requirements may be significantly lower than traditional, high reliability communication demands, the proposed method is better suited to find scheduling configurations that can moreover meet the strict latency requirements imposed by the physical system. This is in contrast to the control-aware approach taken in \cite{GatsisEtal15}, which focuses on low-power and infrequent transmissions rather than the low-latency setting of interest in industrial control.

The paper is organized as follows. We formulate the wireless control system in which state information is communicated to the control over a wireless channel. Due to the potential for random packet drops, this is modeled as a switched dynamical system (Section \ref{sec_problem_formulation}). A Lyapunov function is used to evaluate the stability of the control state, and the uncertainty in this measurement grows the more consecutive packets are lost for a particular system.  We then discuss the scheduling parameters of the IEEE 802.11ax communication model (Section \ref{sec_comm_model}).
From there, we derive a mathematical formulation of the optimal scheduling problem (Section \ref{sec_optimal}). This can be formulation by minimizing a control cost with an explicitly latency constraint (Section \ref{sec_optimal_a}) or minimizing transmission time with an explicit control performance constraint (Section \ref{sec_optimal_b}).

Using this formulation, we develop the control-aware low latency scheduling (CALLS) method (Section \ref{sec_calls}). The CALLS method uses current control states and channel conditions to derive dynamic packet success rates for each user (Section \ref{sec_psr}). In this way, control systems that are closest to instability will be given priority in the scheduling so that they may close their control loops. The scheduling procedure consists of a random user selection procedure to reduce the number of required PPDUs that incur significant overhead  (Section \ref{sec_rss}), followed by an assignment-method based scheduling of selected users to minimize total transmission time (Section \ref{sec_assignment}). The performance of the CALLS method is analyzed in a series of simulation experiments in which its performance is compared against a control-agnostic procedure (Section \ref{sec_simulation}). We demonstrate in numerous low-latency control systems that the control-aware method can support more users than the alternative and achieve more robust overall performance. 

\section{Wireless Control Sysyem}\label{sec_problem_formulation}
\begin{figure}
\centering
\pgfdeclarelayer{bg0}    
\pgfdeclarelayer{bg1}    
\pgfsetlayers{bg0,bg1,main}  

\tikzstyle{block} = [draw,rectangle,thick,
text height=0.2cm, text width=0.7cm, 
fill=blue!30, outer sep=0pt, inner sep=0pt]
\tikzstyle{dots} = [font = \large, minimum width=2pt]
\tikzstyle{dash_block} = [draw,rectangle,dashed,minimum height=1cm,minimum width=1cm]
\tikzstyle{smallblock} = [draw,rectangle,minimum height=0.5cm,minimum width=0.5cm,fill= green!30, font =  \scriptsize]
\tikzstyle{smallcircle} = [draw,ellipse,minimum height=0.1cm,minimum width=0.3cm,fill= yellow!40, font =  \scriptsize ]
\tikzstyle{connector} = [->]
\tikzstyle{dash_connector} = [->,thick,decorate,decoration={snake, amplitude =1pt, segment length=8pt}, magenta]
\tikzstyle{branch} = [circle,inner sep=0pt,minimum size=1mm,fill=black,draw=black]

\tikzstyle{vecArrow} = [thick, decoration={markings,mark=at position
   1 with {\arrow[semithick]{open triangle 60}}},
   double distance=1.4pt, shorten >= 5.5pt,
   preaction = {decorate},
   postaction = {draw,line width=1.4pt, white,shorten >= 4.5pt}]

\begin{tikzpicture}[scale=1, blocka/.style ={rectangle,text width=0.9cm,text height=0.6cm, outer sep=0pt}]
 \small

    \matrix(M)[ampersand replacement=\&, row sep=2.0cm, column sep=10pt] {
    
    \node[smallblock, align=center] (CS1) {Control \\ System {1}};\&\&
    \node[smallblock, align=center] (CS2) {Control \\ System {2}};\&\&\&
    \node(d1) {$\cdots$};\&
    \node[smallblock, align=center] (CSm) {Control \\ System \textit{m}};\&
    \\
    \node[blocka] (R1) {};\&\&
    \node[blocka] (R2) {};\&\&\&
    \node[blocka] (d3) {};\&
    \node[blocka] (Rm) {};\&
    \\
    };

    \node[block] (outer) [fit=(R1.north west) (d3) (Rm.south east)] {};
    
    \node[align=center, scale =0.9] at (outer.center) {Access Point/ \\Controller};
    
    \draw [->, thick, red] (CS1) -- node[left]{} (R1);
    \draw [->, thick, red] (CS2) -- node[left]{} (R2);
    \draw [->, thick, red] (CSm) -- node[left]{} (Rm);

		\begin{pgfonlayer}{bg0}    
		\draw [->, dashed, black] (R1) |- ($(R1) + (+35pt,-20pt)$) node(down_right){} 
		-- ($(CS1) + (+35pt,+20pt)$) node(up_right){} -| (CS1);
		\end{pgfonlayer}

		\begin{pgfonlayer}{bg0}    
		\draw [->, dashed, black] (R2) |- ($(R2) + (+35pt,-20pt)$) node(down_right){} 
		-- ($(CS2) + (+35pt,+20pt)$) node(up_right){} -| (CS2);
		\end{pgfonlayer}

		\begin{pgfonlayer}{bg0}
		\draw [->, dashed, black] (Rm) |- ($(Rm) + (+35pt,-20pt)$) node(down_right){} 
		--($(CSm) + (+35pt,+20pt)$) node(up_right){} -| (CSm);
		\end{pgfonlayer}

		\begin{pgfonlayer}{bg1}
		\node(shared) [fill=red!10, fit={($(CS1.south) + (-15pt, -10pt)$) 
		($(CS2.south) + (-10pt, -10pt)$)
		($(CSm.south) + (+20pt, -10pt)$)
		($(R1.north) + (-15pt, +10pt)$)
		($(R2.north) + (-10pt, +10pt)$)
		($(Rm.north) + (+20pt, +10pt)$)
		}] {};
		\end{pgfonlayer}
		
		\node[align=center, red!50](shared_medium) at (shared.center) {Shared \\ Wireless \\ Medium};

\coordinate (FIRST NE) at (current bounding box.north east);
   \coordinate (FIRST SW) at (current bounding box.south west);

	\pgfresetboundingbox
   \useasboundingbox ($(FIRST SW) + (+30pt,0)$) rectangle (FIRST NE);

\end{tikzpicture}
\caption{Wireless control system with $m$ independent systems. Each system contains a sensor that measure state information, which is transmitted to the controller over a wireless channel. The state information is used by the controller to determine control policies for each of the systems.The communication is assumed to be wireless in the uplink and ideal in the downlink.}
\label{fig_wcs}
\end{figure}
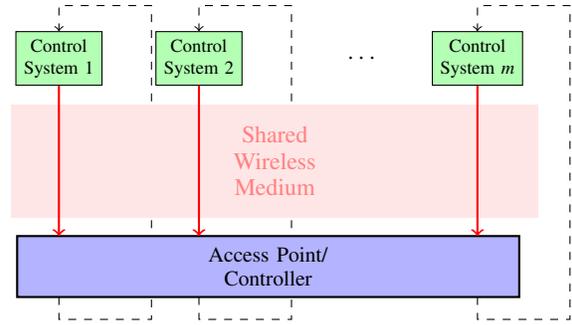

Consider a system of $m$ independent linear control systems, or devices, where each system $i=1,\hdots,m$ maintains a state variable $\bbx_{i} \in \reals^p$. The dynamics are discretized so that the state evolves over time index $k$.  Applying an input $\bbu_{i,k} \in \reals^q$ causes the state and output to evolve based on the discrete-time  state space equations,
\begin{align}\label{eq_control_orig}
\bbx_{i,k+1} &= \bbA_{i} \bbx_{i,k} + \bbB_{i} \bbu_{i,k} + \bbw_k
\end{align}
where $\bbA_{i} \in \reals^{p \times p}$ and $\bbB_{i} \in \reals^{p \times q}$ are matrices that define the system dynamics, and $\bbw_{k} \in \reals^{p}$ is Gaussian noise with co-variance $\bbW_i$ that captures the errors in the linear model (due to, e.g., unknown dynamics or from linearizion of non-linear dynamics). We further assume the state transition matrix $\bbA_i$ is on its own unstable, i.e. has at least one eigenvalue greater than 1. This is to say that, without an input, the dynamics will drive the state $\bbx_{i,k} \rightarrow \infty$ as $k \rightarrow \infty$.

In the wireless control system model is presented in Figure \ref{fig_wcs}. Each system is closed over a wireless medium, over which the sensor located at the control system sends state information to the controller located at a wireless access point (AP) shared among all systems. Using the state information $\bbx_{i,k}$ received from device $i$ at time $k$, the controller determines the input $ \bbu_{i,k}$ to be applied. We stress in Figure \ref{fig_wcs} we restrict our attention to the wireless communications at the sensing, or ``uplink'', while the control actuation, or ``downlink, is assumed to occur over an ideal channel. We point out that while a more complete model may include packet drops in the downlink, in practice the more significant latency overhead occurs in the uplink. We therefore keep this simpler model for mathematical coherence. In low-latency applications, a high state sampling rate is required be able to adapt to the fast-moving dynamics  This subsequently places a tight restriction on the latency in the wireless transmission, so as to avoid losing sampled state information. This specific latency requirement between the sensor and AP we denote by $\tau_{\max}$, and is often considered to be in the order of milliseconds.

Because the control loop in Figure \ref{fig_wcs} is closed over a wireless channel, there exists a possibility at each cycle $k$ that the transmission fails and state information is not received by the controller. We refer to this as the ``open-loop' configuration; when state information is received, the system operates in ``closed-loop.'' As such, it is necessary to define the system dynamics in both configurations. Consider a generic linear control, in which the input being determined as $ \bbu_{i,k} = \bbK_i \bbx_{i,k}$ for some matrix $\bbK_i \in \reals^{q \times p}$. Many common control policies indeed can be formulated in such a manner, such as LQR control. In general, this matrix $\bbK$ is chosen such as that the closed loop dynamic matrix $\bbA + \bbB \bbK$ is stable, i.e. has all eigenvalues less that 1. Thus, application of this control over time will drive the state $\bbx_{i,k} \rightarrow 0$ as $k \rightarrow \infty$. As the controller does not always have access to state information, we alternatively consider the estimate of state information of device $i$ known to the controller at time $k$ as
\begin{align}\label{eq_state_est}
\hbx^{(l_i)}_{i,k} :=(\bbA_i + \bbB_i\bbK_i)^{l_i} \bbx_{i,k-l_i},
\end{align}
where $k-l_i \geq k-1$ is the last time instance in which control system $i$ was closed. There are two important things to note in \eqref{eq_state_est}. First, this is the estimated state \emph{before} a transmission has been attempted at time $k$; hence, $l_i = 1$ when state information was received at the previous time. Second, observe that in \eqref{eq_state_est} we assume that the AP/controller has knowledge of the dynamics $\bbA_i$ and $\bbB_i$, as well as the linear control matrix $\bbK_i$. Any gap in this knowledge of dynamics is captured in the noise $\bbw_k$ in the actual dynamics in \eqref{eq_control_orig}. Note that the estimated state \eqref{eq_state_est} is used in place of the true state in both the determination of the control \emph{and} the radio resource allocation decisions as discussed later in this paper.

At time $k$, if the state information is received, the controller can apply the input $\bbu_{i,k} = \bbK_i \bbx_{i,k}$ exactly, otherwise it applies an input using the estimated state, i.e. $\bbu_{i,k} = \bbK_i \hbx_{i,k}$. Thus, in place of \eqref{eq_control_orig}, we obtain the following switched system dynamics for $\bbx_{i,k}$ as
\begin{align}\label{eq_control_switch}
\bbx_{i,k+1} &= \begin{cases}
(\bbA_i + \bbB_i \bbK_i) \bbx_{i,k} + \bbw_k, \ \text{in closed-loop}, \\
\bbA_i \bbx_{i,k} + \bbB_i\bbK_i\hbx^{(l_i)}_{i,k} + \bbw_k, \ \text{in open-loop}.
\end{cases}
\end{align}
The transmission counter $l_i$ is updated at time $k$ as
\begin{align}
l_i &\leftarrow \begin{cases}
1, \ \text{in closed-loop}, \\
l_1 + 1, \ \text{in open-loop}.
\end{cases} \label{eq_time_switch}
\end{align}
Observe in \eqref{eq_control_switch} that, when the system operates in open loop, the control is not applied relative to the current state $\bbx_{i,t}$ but on the estimated state $\hbx^{(l_i)}_{i,k}$, which indeed may not be close to the true state. In this case, the state may not be driven to zero as in the closed-loop configuration. To see the effect of operating in open loop for many successive iterations, we can write the error between the true and estimated state as
\begin{align}\label{eq_diff}
\bbe_{i,k} := \bbx_{i,k} - \hbx^{(l_i)}_{i,k} = \sum_{j=0}^{l_i-1}\bbA_i^{j} \bbw_{i,k-j-1}.
\end{align}
In \eqref{eq_diff}, it can be seen that as $l_i$ grows, the error $\bbe_{i,k}$ grows with the accumulation of the noise present in the actual state but not considered in the estimated state. Thus, if $l_i$ is large and $\bbw_{i,k}$ is large (i.e., high variance), this error will become large as well.

To conclude the development of the wireless control formulation, we define a quadratic Lyapunov function $L(\bbx) := \bbx^T \bbP \bbx$ for some positive definite $\bbP \in \reals^{p \times p}$ that measures the performance of the system as a function of the state. Because the scheduler only has access to estimated state info, we consider the expected value of the Lagrangian given the state estimate, which can be found via \eqref{eq_diff} as
\begin{align}
\E [L(\bbx_{i,k}) \mid &\hbx^{(l_i)}_{i,k}]  \label{eq_perf} \\
&=  (\hbx^{(l_i)}_{i,k})^T \bbP (\hbx^{(l_i)}_{i,k}) + \sum_{j=0}^{l_i-1} \Tr[(\bbA_i^T\bbP^{\frac{1}{j}}\bbA_i)^{j} \bbW_i]. \nonumber
\end{align}
Thus, the control-specific goal is to keep $\E [L(\bbx_{i,k}) \mid \hbx^{(l_i)}_{i,k}]$ within acceptable bounds for each system $i$. We now proceed to discuss the wireless communication model that determines the resource allocations necessary to close the loop.

\subsection{IEEE 802.11ax communication model}\label{sec_comm_model}

\begin{figure}
\includegraphics[width=0.5\textwidth]{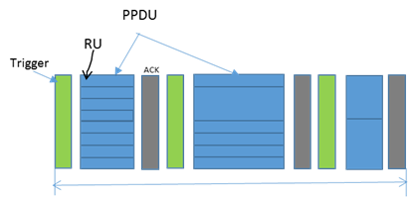}
\caption{Multiplexing of frequencies (RU) and time (PPDU) in IEEE 802.11ax transmission window (formally referred as Transmission Opportunity or TXOP in the standard. The total transmission time is the time of all PPDUs, including the overhead of trigger frames (TF) and acknowledgments.}
\label{fig_multiplex}
\end{figure}

We consider the communication model provided in the next-generation Wi-Fi standard IEEE 802.11ax. While 3GPP wireless systems such as LTE~\cite{sesia2011lte} or the next generation 5G~\cite{agiwal2016next} can also be considered as alternate communication models, most factory floors are already equipped with Wi-Fi connectivity and, moreover, Wi-Fi can operate in the unlicensed band. It is generally considered to be cost-effective to operate and maintain.

Traditional Wi-Fi systems rely only on contention-based channel access and may introduce high or variable latency in congested or dense deployment scenarios even in a fully managed Wi-Fi network, which is typically available in industrial control and automation scenarios. To address the problems with dense deployment, the draft 802.11ax amendment has defined scheduling capability for Wi-Fi access points (APs). Wi-Fi devices can now be scheduled for accessing the channel in addition to the traditional contention-based channel access. Such scheduled access enables more controlled and deterministic behavior in the Wi-Fi networks. Within each transmission window (formally referred as transmission opportunity or TXOP in the standard), the AP may schedule devices through both frequency and time division multiplexing using the multi-user (MU) OFDMA technique. This to say that devices can be slotted in various frequency bands---formally called resource units (RUs)---and in different timed transmission slots---formally called PPDUs. An example of the multiplexing of devices across time and frequency is demonstrated in Figure \ref{fig_multiplex}. The AP additionally sends a trigger frame (TF) indicating which devices should transmit data in the current TXOP and the time/frequency resources these triggered devices should use in their transmissions. 

To state this model formally, the scheduling parameters assigned by the AP to each device consist of a frequency-slotted RU, time-slotted PPDU, and an associated modulation and coding scheme (MCS) to determine the transmission format. The transmission power is assumed to be fixed and equally divided amongst all devices. We define the following notations to formulate these parameters.  To specify an RU, we first notate by $f^1, f^2, \hdots, f^b$, where $n$ is the number of discrete frequency bands of fixed bandwidth (typically 2 MHz) in which a device can transmit; in a 20MHz channel, for example, there are $n=10$ such bands. For each device, we then define a set of binary variables $\varsigma^j_{i} \in \{0,1\}$ if device $i$ transmits in band $f^j$ and collect all such variables for device $i$ in $\bbsigma_i = [\varsigma^1_{i};\hdots;\varsigma^b_{i}] \in \{0,1\}^{b}$ and for all devices in  $\bbSigma := [\bbsigma_i, \hdots, \bbsigma_m] \in \{0,1\}^{b \times m}$. A device may transmit in bands in certain multiples of 2MHz as well, which would be notated as, e.g. $\bbsigma_i = [1; 1; 0; \hdots; 0]$ for transmission in an RU of size 4MHz. Note, however, that allowable RU's contain only sizes of certain multiples of 2MHz---namely, 2MHz, 4MHz, 8MHz, and 20MHz in the 802.11ax standard. Furthermore, it is only permissible to transmit in \emph{adjacent} bands, e.g. $f^{j}$ and $f^{j+1}$. We therefore define the set $\ccalS \subset \{0,1\}^{b}$ as the set of binary vectors that define permissible RUs and consider only $\bbsigma_i \in \ccalS$ for all devices $i$.  Finally, note that the RU assignment $\bb0 \in \ccalS$ signifies a device does not transmit in this particular transmission window.

\begin{table}
\begin{tabular}{ l | l | l | l  }
\hline
  $\mu$ & Modulation type & Coding rate & Data rate (Mb/s)  \\ \hline
  0 & BPSK & 1/2 & 4 \\
  1 & QPSK & 1/2 & 16 \\
  2 & QPSK & 3/4 & 24 \\
  3 & 16-QAM & 1/2 & 33 \\
  4 & 16-QAM & 3/4 & 49 \\
  5 & 64-QAM & 2/3 & 65 \\
  6 & 64-QAM & 3/4 & 73 \\
  7 & 64-QAM & 5/6 & 81 \\
  8 & 256-QAM & 3/4 & 98 \\
  9 & 256-QAM & 5/6 & 108 \\
  10 & 1024-QAM & 3/4 & 122 \\
  \hline  
\end{tabular}
\caption{Data rates for MCS configurations in IEEE 802.11ax for 20MHz channel. The modulation type and coding rate in the first 2 columns together specify a PDR function $q(\bbmu,\bbsigma)$ for RU $\bbsigma$. The data rate in the third column specifies the associated transmission time $\tau(\mu, \bbsigma)$. }
\label{tab_mcs}
\end{table}

To specify the PPDU of all scheduled devices, we define for device $i$ a positive integer value $\alpha_i \in \mathbb{Z}_{++}$ that denotes the PPDU slot in which it transmits and collect such variables for all devices in $\bbalpha = [\alpha_1; \hdots; \alpha_m] \in \mathbb{Z}_{++}^m$. Likewise, device $i$ is given an MCS $\mu_i$ from the discrete space $\ccalM = \{0,1,2,\hdots,10\}$. The MCS in particular defines a pair of modulation scheme and coding rate that subsequently determine both the data rate and packet error rate of the transmission. The allowable MCS settings provided in 802.11ax are provided in Table \ref{tab_mcs}. Finally, we notate by $\bbh_{i} := [h_i^1; h_i^2; \hdots; h_i^b] \in \reals^b_+$ a set of channel states experienced by device $i$, where $h_i^j$ is the gain of a wireless fading channel in frequency band $f^j$. We assume that channel conditions are constant within a single TXOP, i.e. do not vary across PPDUs.

We now proceed to define two functions that describe the wireless communications over the channel. Firstly, we define a function  $q: \reals_+^b \times \ccalM \times \ccalS \rightarrow [0,1]$ which, given a set of channel conditions $\bbh$, MCS $\mu$, and RU $\bbsigma$, returns the probability of successful transmission, otherwise called packet delivery rate (PDR). Furthermore, define by $\tau: \ccalM \times \ccalS  \rightarrow \reals_+$ a function that, given an MCS $\mu$ and RU $\bbsigma$, returns the maximum time taken for a single transmission attempt. Assuming a fixed packet size, such a function can be determined from the data rates associated with each MCS in Table \ref{tab_mcs}.  Observe that all functions just defined are determined independent of the PPDU slot the transmission takes place in, while transmission time is also independent of the channel state. Because a PPDU cannot finish until all transmissions within the PPDU have been completed, the total transmission time of a single PPDU $s$ is the maximum transmission time taken by all devices within that time slot. We define the transmission time of PPDU slot $s$ as
\begin{equation}\label{eq_time_slot}
\hat{\tau}(\bbSigma, \bbmu, \bbalpha, s) := \max_{i: \alpha_i = s} \tau(\mu_i, \bbsigma_i) + \tau_0(\bbalpha,s),
\end{equation}
where $\tau_0: \mathbb{Z}_{++}^m \times \mathbb{Z}_{++} \rightarrow \reals_+$ is a function that specifies the communication overhead of PPDU $s$. This overhead may consist of, e.g., the time required to send TFs to scheduled users, as seen in Figure \ref{fig_multiplex}.


\section{Optimal Control Aware Scheduling}\label{sec_optimal}
Using the communication model of 802.11ax just outlined and the control-based Lyapunov metric of \eqref{eq_perf}, we can formulate an optimization problem that characterizes the exact optimal scheduling of transmissions with a transmission window to maximize control performance. The optimal scheduling and allocation selects the set of RUs $\bbSigma$. MCS $\bbmu$, and PPDUs $\bbalpha$ for all devices--which in effect fully determine the schedule---such to minimize a cost subject to scheduling design and feasibility constraints. In particular, we discuss two related, alternative formulations of the low-latency scheduling problem.

\subsection{Latency-constrained scheduling}\label{sec_optimal_a}
In the latency-constrained formulation, we are interested in minimizing a common control cost subject to strict latency requirements. In particular, in the low-latency setting we set a bound $\tau_{\max}$ on the total transmission time across all PPDUs in a TXOP. This constraint is relevant in design of MAC-layer protocols that set strict limits on transmission times. In addition, the RU and PPDU allocation across devices must be feasible, i.e., two devices cannot be transmitting in the same frequency band in the same PPDU.

Recall the PDR function $q(\bbh,\mu,\bbsigma)$ and consider that this can alternatively be interpreted as the probability of closing the control loop under certain channel conditions and scheduling parameters. From there, we can now write the expected Lyapunov value for at time $k+1$ given its current state $\bbx_{i,k}$, channel state $\bbh_{i,k}$, MCS $\mu_i$, and RU $\bbsigma_i$ using the expected cost in \eqref{eq_perf}. By defining $\bbx_{i,k+1}^c$ and $\bbx_{i,k+1}^o$ as the closed loop and open loop states, respectively, as determined by the switched system in \eqref{eq_control_switch}, this is written as
\begin{align}
J_i(\hbx^{(l_i)}_{i,k},\bbh_{i,k},\mu_i,\bbsigma_i) &:= \E (L(\bbx_{i,k+1}) \mid \hbx^{(l_i)}_{i,k},\bbh_{i,k},\mu_i,\bbsigma_i) \nonumber \\
=& (1-q(\bbh_{i,k},\mu_i,\bbsigma_i))  \E L(\bbx_{i,k+1}^o \mid \hbx^{(l_i)}_{i,k}) \nonumber \\
&+q(\bbh_{i,k},\mu_i,\bbsigma_i) \E L(\bbx_{i,k+1}^c \mid \hbx^{(l_i)}_{i,k}). \label{eq_lyap_p}
\end{align}

For notational convenience, we collect all current estimated control states at time $k$ as $\hbX_k := [\hbx^{(l_1)}_{1,k}, \hdots, \hbx^{(l_m)}_{m,k}]$ and channel states $\bbH_k = [\bbh_{1,k}, \hdots, \bbh_{m,k}]$. Now, define the total control cost, given the current states and scheduling parameters as some aggregation of the combined expected future Lyapunov costs across all devices, i.e.,
\begin{align}\label{eq_cost_total}
 \tdJ( \hbX_{k},&\bbH_{k}, \bbmu, \bbSigma) := \\
 &g(J_1(\hbx^{(l_1)}_{1,k},\bbh_{1,k},\mu_1,\bbsigma_1), \hdots, J_m(\hbx^{(l_m)}_{m,k},\bbh_{m,k},\mu_m,\bbsigma_m)). \nonumber
 \end{align}
 Natural choices of the aggregation function $g(\cdot)$ are, for example, either the sum or maximum of its arguments. 

The optimal scheduling at transmission time $k$ is formulated as the one which minimizes this cost $\tdJ$ while satisfying low-latency and feasibility requirements of the schedule, expressed formally with the following optimization problem.
\begin{align}\label{eq_problem}
[&\bbSigma_k^*, \bbmu_k^*,\bbalpha_k^*]  := \argmin_{\bbSigma, \bbmu, \bbalpha, S} \tdJ(\hbX_{k},\bbH_{k}, \bbmu, \bbSigma) \\
& \quad \st \sum_{i: \alpha_i = s} \varsigma_{i}^j \leq 1, \quad \forall j, s, \label{eq_c1} \\
& \qquad \quad \sum_{s=1}^S \hat{\tau}(\bbSigma, \bbmu, \bbalpha, s)  \leq \tau_{\max}, \label{eq_c2}\\
& \qquad \quad 1 \leq \alpha_i \leq S, \quad \forall i, \label{eq_c3}\\
& \qquad \quad \bbsigma_i \in \ccalS, \ \forall i, \quad \bbmu \in \ccalM^m, \quad \bbalpha \in \mathbb{Z}_+^m, \quad S \in \mathbb{Z}_+. \label{eq_c4}
\end{align}
The optimization problem in \eqref{eq_problem} provides a precise and instantaneous selection of frequency allocations between devices given their current control states $\hbX_k$ and communication states $\bbH_k$. The constraints in \eqref{eq_c1}-\eqref{eq_c4} encode the following scheduling conditions. The constraint \eqref{eq_c1} ensures that for every PPDU $s$, there is only one device transmitting on a frequency slot $j$. In \eqref{eq_c2}, we set the low-latency transmission time constraint in terms of the sum of all transmission times for each PPDU $s$. The constraint in \eqref{eq_c3} bounds each transmission slot by the total number of PPDU's $S$ while \eqref{eq_c4} constrains each variable to its respective feasible set. Note that $S$ is itself treated as an optimization variable in the above problem, so that the number of PPDUs may vary as needed.

Observe in the objective in \eqref{eq_problem} that, by minimizing an aggregate of local control costs, the devices with the highest cost $J_i$ as described by \eqref{eq_lyap_p} will be given the most bandwidth or most favorable frequency bands to increase probability of successful transmission $q(\bbh_{i,k},\mu_i,\bbsigma_i)$. This in effect increases the chances those devices will close their control loops and be driven towards a more favorable state. Likewise, a device who is experiencing very adverse channel conditions may not be allocated prime transmission slots to reserve such resources who have more favorable channel conditions. In this way, we say this is \emph{control-aware} scheduling, as it considers both the control and channel states of the devices to determine optimal scheduling. However, we stress that the optimization problem described in \eqref{eq_problem}-\eqref{eq_c4} is by no means easy to solve. In fact, the optimization over multiple discrete variables makes this problem combinatorial in nature. In the following section, we discuss a practical reformulation of the problem above and develop heuristic methods to approximate the solutions in realistic low-latency wireless applications.


\subsection{Control-constrained scheduling}\label{sec_optimal_b}
We reformulate the problem in \eqref{eq_problem}-\eqref{eq_c4} to an alternative formulation that more directly informs the control-aware, low-latency scheduling method to be developed. To do so, we introduce a \emph{control-constrained} formulation, in which the Lyapunov decrease goals are presented as explicit requirement, i.e. constraints in the optimization problem. We are interested, then, in constraint of the form
\begin{align}\label{eq_control_constraint}
\tdJ(\hbX_{k},\bbH_{k}, \bbmu, \bbSigma) \leq J_{\max},
\end{align}
where $J_{\max}$ is a limiting term design to enforce desired system performance. Determining this constant is largely dependent on the particular application of interest, needs of the control systems, and also may be related to the choice aggregation function $g(\cdot)$ in \eqref{eq_cost_total}. For example, $J_{\max}$ may represent a point at which control systems become volatile, unsafe, or unstable.

For the scheduling procedure developed in this paper, we focus on a particular formulation of the control constraint in \eqref{eq_control_constraint} that constrains the expected future Lyapunov value of each system by a rate decrease of its current value. In particular the following rate-decrease condition for each device $i$,
\begin{align}\label{eq_lyap_constraint}
J_i(\hbx^{(l_i)}_{i,k},\bbh_{i,k},\mu_i,\bbsigma_i) \leq \rho_i \E [L(\bbx_{i,k}) \mid \hbx^{(l_i)}_{i,k}] + c_i,
\end{align}
where $\rho_i \in (0,1]$ is a decrease rate and $c_i \geq 0$ is a constant. Recall the definition of $J_i(\hbx^{(l_i)}_{i,k},\bbh_{i,k},\mu_i,\bbsigma_i)$ in \eqref{eq_lyap_p} as the expected Lyapunov value of time $k+1$ given its current estimate and scheduling $\mu_i, \bbsigma_i$. The constraint in \eqref{eq_lyap_constraint} ensures the future Lyapunov cost will exhibit a decrease of at least a rate of $\rho_i$ for device $i$ in expectation. The constant $c_i$ is included to ensure this condition is satisfied by default if the state $\hbx^{(l_i)}_{i,k}$ is already sufficiently small. 

We formulate the control-constrained scheduling problem by substituting the latency constraint with the control constraint in \eqref{eq_lyap_constraint}, i.e.,
\begin{align}\label{eq_problem2}
[&\bbSigma_k^*, \bbmu_k^*,\bbalpha_k^*]  := \argmin_{\bbSigma, \bbmu, \bbalpha, S} \sum_{s=1}^S \hat{\tau}(\bbSigma, \bbmu, \bbalpha, s) \\
& \quad \st \sum_{i: \alpha_i = s} \varsigma_{i}^j \leq 1, \quad \forall j, s, \label{eq_c12} \\
& \quad   J_i(\hbx^{(l_i)}_{i,k},\bbh_{i,k},\mu_i,\bbsigma_i) \leq \rho \E [L(\bbx_{i,k}) \mid \hbx^{(l_i)}_{i,k}] + c_i \ \forall i, \label{eq_c22}\\
& \qquad \quad 1 \leq \alpha_i \leq S, \quad \forall i, \label{eq_c32}\\
& \qquad \quad \bbsigma_i \in \ccalS, \ \forall i, \quad \bbmu \in \ccalM^m, \quad \bbalpha \in \mathbb{Z}_+^m, \quad S \in \mathbb{Z}_+. \label{eq_c42}
\end{align}
Observe that the objective in \eqref{eq_problem2} is now to minimize the total transmission time, rather than being forced as an explicit constraint. In this way, the optimization problem defined in 
\eqref{eq_problem2}-\eqref{eq_c42} can be viewed as an alternative to the latency constrained problem in \eqref{eq_problem}-\eqref{eq_c4}. Because the scheduling algorithm we develop in this paper requires the ability to quickly identify feasible solutions, we focus our attention on the control-constrained formulation in \eqref{eq_problem2}-\eqref{eq_c42}. Before presenting the details of the scheduling algorithm, we present a brief remark regarding the  addition of ``safety'', or worst-case, constraints to either problem formulation.

\begin{remark}\label{remark_worst_case}\normalfont
The control constraint in \eqref{eq_c22} is formulated to guarantee an average decrease of expected Lyapunov value by a rate of $\rho$. This is of interest to ensure the system states are driven to zero over time. However, in practical systems we may also be interested in protecting against worst-case behavior, e.g. entering an unsafe or unstable region. Consider a vector $\bbb_i \in \reals^p$ as the boundary of safe operation of system $i$. A constraint that protects against exceeding this boundary can be written as
\begin{align}\label{eq_worst_case}
\P [ |\bbx_{i,k+1}| \geq \bbb_i \mid \hbx^{(l_i)}_{i,k},\bbh_{i,k},\mu_i,\bbsigma_i ] \leq \delta,
\end{align}
where $\delta \in (0,1)$ is small. The expression in \eqref{eq_worst_case} can be included as an additional constraint to either the latency-constrained or control-constrained scheduling problems previously discussed.
\end{remark}

\section{Control-Aware Low-Latency Scheduling (CALLS)}\label{sec_calls}
We develop a control-aware low-latency scheduling (CALLS) algorithm to approximately solve the control-constrained scheduling formulation in \eqref{eq_problem2}-\eqref{eq_c42}. Because this problem is combinatorial in nature, it is infeasible to solve exactly. Instead, we focus on a practical and efficient means of solving approximately. In particular, we identify sets of feasible points and use a heuristic approach towards minimizing the transmission time objective among the set of feasible points. Additionally, within the development of the CALLS method we identify and characterize new PDR requirements that are defined relative to the control system requirements; these are generally significantly less strict than the PDR requirements often considered in general high reliability communication systems without codesign. Overall, the CALLS method consists of (i) the derivation of dynamic control-based PDR targets, (ii) a principled random selection of devices to schedule to reduce latency, and (iii) the use of assignment based methods to find a low-latency schedule. We discuss these three components in detail in the proceeding subsections. 

\subsection{Control-based dynamic PDR}\label{sec_psr}
Due to the complexity of the scheduling problem in \eqref{eq_problem2}-\eqref{eq_c42}, we first focus our attention on identifying scheduling parameters $\{\bbSigma_k, \bbmu_k, \bbalpha_k\}$ that are feasible, i.e. satisfy the constraints in \eqref{eq_c12}-\eqref{eq_c42}. In particular, the Lyapunov control constraint in \eqref{eq_c22} is of significant interest. Recall that the control cost function $J_i(\hbx^{(l_i)}_{i,k},\bbh_{i,k},\mu_i,\bbsigma_i)$ is itself determined by the PDR $q(\bbh_{i,k},\mu_i,\bbsigma_i)$, as per \eqref{eq_lyap_p}. Thus, the constraint in \eqref{eq_c22} can be seen as indirectly placing a constraint on the required PDR necessary to achieve a $\rho_i$-rate decrease in expectation. The equivalent condition on PDR $q(\bbh_{i,k},\mu_i,\bbsigma_i)$ is presented in the following proposition.

\begin{proposition}\label{prop_pdr_constraint}
Consider the Lyapunov control constraint in \eqref{eq_c22} and the definition of $J_i(\hbx^{(l_i)}_{i,k},\bbh_{i,k},\mu_i,\bbsigma_i)$ given in \eqref{eq_lyap_p}.  Define the closed-loop state transition matrix $\bbA^c_i := \bbA_i + \bbB_i \bbK_i$ and $j$-accumulated noise $\omega_i^j := \Tr[(\bbA_i^T\bbP^{1/j} \bbA_i)^{j} \bbW_i]$. The control constraint in \eqref{eq_c22} is satisfied for device $i$ if and only if the following condition on PDR $q(\bbh_{i,k},\mu_i,\bbsigma_i)$ holds, 
\begin{align}\label{eq_pdr_constraint}
&q(\bbh_{i,k},\mu_i,\bbsigma_i) \geq \tdq_i(\hbx^{(l_i)}_{i,k}) :=  \\ &\ \frac{1}{\Delta_i} \left[  \left\| (\bbA^c_i - \rho_i\bbI)\hbx^{(l_i)}_{i,k} \right\|_{\bbP^{\frac{1}{2}}}^2+ (1-\rho_i)\sum_{j=0}^{l_i-1} \omega_i^j + \omega_i^{l_i} - c_i  \right],  \nonumber
\end{align}
where we have further defined the constant
\begin{equation}\label{eq_noise_diff}
\Delta_i :=  \sum_{j=0}^{l_i-1}[\omega^{j+1}_i - \Tr(\bbA^{cT}_i (\bbA_i^T\bbP^{1/j}\bbA_i)^j \bbA^c_i\bbW_i)  ].
\end{equation}
\end{proposition}
\begin{myproof}
 Consider the Lyapunov decrease constraint as written in \eqref{eq_c22}. As the same logic holds for all $i$ and $k$, for ease of presentation we remove all subscripts when presenting the details of this proof. We further introduce the simpler notation $q := q(\bbh,\mu,\bbsigma)$. Now, we may expand the left hand side of \eqref{eq_c22} be rewriting the definition in \eqref{eq_lyap_p} as
\begin{align}
J(\hbx^{(l)},\bbh,\mu,\bbsigma)  &= q \E_{\bbw}[ L(\bbA_c \bbx + \bbw)] \label{eq_p1}\\
&\quad + (1-q)  \E_{\bbw} [L(\bbA \bbx + \bbB\bbK\hbx + \bbw)]. \nonumber 
\end{align}
Recall the definition of the quadratic Lyapunov function $L(\bbx) := \bbx^T \bbP \bbx$ for some positive definite $\bbP$. Further recall the relation $\bbx = \hbx + \bbe$ as described by \eqref{eq_diff}. Combining these, we expand the right hand size of \eqref{eq_p1} as
\begin{align}
&J(\hbx^{(l)},\bbh,\mu,\bbsigma) = \label{eq_p2}\\
&\quad q \E_{\bbw} \left[ \bbA_c  \left(\hbx +\bbe \right) + \bbw \right]^T \bbP \left[ \bbA_c  \left(\hbx +\bbe \right) + \bbw \right]  \nonumber\\
&\ +  (1-q) \E_{\bbw} \left[ \bbA_c \hbx + \bbA\bbe +  \bbw \right]^T \bbP \left[ \bbA_c \hbx +\bbA\bbe +  \bbw\right]. \nonumber
\end{align}
To evaluate the expectations in \eqref{eq_p2}, recall the random noise $\bbw$ follows a Gaussian distribution with zero mean and covariance $\bbW$. Thus, the expectation can be evaluated over $\bbw$ and expanded as
\begin{align}
&J(\hbx^{(l)},\bbh,\mu,\bbsigma)= \\
&q \left[  \| \bbA_c \hbx\|^2_{\bbP^{\frac{1}{2}}} + \Tr(\bbP\bbW) + \sum_{j=0}^{l-1} \Tr( \bbA_c (\bbA^T\bbP^{\frac{1}{j}} \bbA)^j \bbA_c \bbW)\right] +  \nonumber\\
&   (1-q)\left[   \| \bbA_c \hbx\|^2_{\bbP^{\frac{1}{2}}} + \Tr(\bbP \bbW) + \sum_{j=1}^{l} \Tr((\bbA^T\bbP^{\frac{1}{j}} \bbA)^{j}\bbW) \right].\nonumber
\end{align}
From here, we rearrange terms and substitute the notation $\omega^j := \Tr[(\bbA^T\bbP^{1/j}\bbA)^{j} \bbW]$ to obtain that the control cost can be written as
\begin{align}
J(\hbx^{(l)},\bbh,\mu,\bbsigma)&=\left[  \| \bbA_c \hbx\|^2_{\bbP^{\frac{1}{2}}}+ \Tr(\bbP \bbW) + \sum_{j=1}^{l}\omega^{j} \right]  \label{eq_p4} \\
&\ + q  \sum_{j=0}^{l-1}[ \Tr( \bbA_c (\bbA^T\bbP^{\frac{1}{j}}\bbA)^j \bbA_c \bbW) - \omega^{j+1}]. \nonumber
\end{align}
With \eqref{eq_p4}, we have expanded the control cost in terms of the PDR $q$. Now, we return to the constraint in \eqref{eq_c22}. Recall the expansion for $\E [L(\bbx) \mid \hbx^{(l)}]$ via \eqref{eq_perf}. By combining this with the expansion in \eqref{eq_p4}, the terms in\eqref{eq_c22} can be rearranged to obtain the inequality in \eqref{eq_pdr_constraint}.
\end{myproof}

In Proposition \ref{prop_pdr_constraint} we establish a lower bound $\tdq_i(\hbx^{(l_i)}_{i,k})$ on the PDR of device $i$ that is dependent upon the current estimated state $\hbx^{(l_i)}_{i,k}$ and system dynamics determined by $\bbA^c_i, \bbA_i$, and $\bbW^i$. We may note the following intuitions about the constraint in \eqref{eq_pdr_constraint}. The PDR condition naturally grows stricter as the bound $\tdq_i(\hbx^{(l_i)}_{i,k})$ defined on the right hand side of \eqref{eq_pdr_constraint} gets larger. The first term on the right hand side reflects the current estimated channel state, and will become larger as the state gets larger. Similarly, the latter two terms on the right hand side together reflect the size of the noise that has accumulated by operating in open loop. When the noise variance $\bbW_i$ is high and when the last-update counter $l_i$ is large, these latter two noise terms will both be large. Thus, both the current magnitude of the control state and the growing uncertainty from infrequent transmissions together determine how large is the PDR requirement in \eqref{eq_pdr_constraint}.

We stress the value of the PDR condition in \eqref{eq_pdr_constraint} is both in its adaptability to the control system state and dynamics, as well as its identification of precise target delivery rates that are necessary to keep the control systems moving towards stability on average. Depending on the particular system dynamics as described in \eqref{eq_control_orig}, such PDR's may be, and often are considerably more lenient than the default target transmission success rates used in practical wireless systems (e.g. $q = 0.999$). Thus, through \eqref{eq_pdr_constraint} we make a claim that, with knowledge of the control system dynamics and targeted \emph{control performance}, we can effectively soften the targeted \emph{communication performance}---or ``reliability''--- accordingly to something more easily obtained in low-latency constrained systems. 

\begin{remark}\normalfont
It is worthwhile to note that by placing a stricter Lyapunov decrease constraint with smaller  rate $\rho_i$ in \eqref{eq_c22}, then the first term on the right hand side of \eqref{eq_pdr_constraint} also grows larger and increases the necessary PDR. Generally, selecting a smaller $\rho$ will result in a faster convergence to stability but will require stricter communication requirements. In fact, we may use the inherent bound on the probability $q(\bbh_{i,k},\mu_i,\bbsigma_i) \leq 1$ to find a lower bound on the Lyapunov decrease rate $\rho_i$ that can be feasibly obtained based upon current control state and system dynamics. This bound, however, may not be obtainable in practice due to the scheduling constraints. In practice, we select $\rho_i$ to be in the interval $[0.90,0.1)$. 
\end{remark}

\subsection{Selective scheduling}\label{sec_rss}
We now proceed to describe the procedure with which we can find a set of feasible scheduling decisions $\{\bbSigma_k, \bbmu_k, \bbalpha_k\}$. To begin, we first consider a stochastically \emph{selective scheduling} protocol, whereby we do not attempt to schedule every device at each transmission cycle, but instead select a subset to schedule a principled random manner. Define by $\nu_{i,k} \in [0,1]$ the probability that device $i$ is included in the transmission schedule at time $k$ and further recall by $q(\bbh_{i,k},\mu_i,\bbsigma_i)$ to be the packet delivery rate with which it transmits. Then, we may consider the \emph{effective} packet delivery rate $\hat{q}$ as 
\begin{align}\label{eq_effective_pdr}
\hat{q}(\bbh_{i,k},\mu_i,\bbsigma_i) = \nu_{i,k} q(\bbh_{i,k},\mu_i,\bbsigma_i)
\end{align}
Selective scheduling is motivated by the ultimate goal of minimizing total transmit time as described in the objective in \eqref{eq_problem2}. As we consider a large number of total devices $m$, scheduling all such devices will require a larger number of PPDU slots---a maximum of 9 devices can transmit within a single PPDU. Recall in \eqref{eq_time_slot} that each additional PPDU requires unavoidable overhead in $\tau_0$, which in aggregation over multiple PPDUs may become a significant bottleneck in minimizing $\hat{\tau}$ or meeting a strict latency requirement $\tau_{max}$. Thus, by decreasing the amount of scheduled devices, we may decrease the number of total PPDUs and the overhead that is added to the total transmission time. 

Observe that by introducing the term $\nu_i$ to the evaluation of effective PDR $\tdq_i$ in \eqref{eq_effective_pdr}, we would thus need to transmit with higher PDR $q(\bbh_{i,k},\mu_i,\bbsigma_i) \geq \tdq_i(\hbx^{(l_i)}_{i,k})/\nu_{i,k}$ to meet the condition in \eqref{eq_pdr_constraint}. While imposing a tighter PDR requirement will indeed require longer transmission times, this added time cost is generally less than the transmission overhead of additional PPDUs. In this work, we use the determine scheduling probability of device $i$ through its PDR requirement $\tdq_i(\hbx^{(l_i)}_{i,k})$ as 
\begin{align}\label{eq_prob_c}
\nu_{i,k} := e^{\tdq_i(\hbx^{(l_i)}_{i,k})-1} .
\end{align}

With \eqref{eq_prob_c}, the probability of scheduling device $i$ increases as the required PDR increases. Notice that, when a transmission is required, i.e. $\tdq_i(\hbx^{(l_i)}_{i,k}) = 1$, then device $i$ is included in the scheduling with probability $1$. In general, devices with very high PDR requirements, e.g. $>0.99$, will be scheduled with very high probability. Thus, the transmission time gains that are provided through selective scheduling using \eqref{eq_prob_c} would be minimal, if non-existent, in high-reliability settings in which PDR requirements remain high at all times. However, with the lower PDR requirement obtained through the control-aware scheduling in \eqref{eq_pdr_constraint}, selective scheduling as the potential to create significant time savings, as will be later shown in Section \ref{sec_simulation} of this paper.

\subsection{Assignment-based scheduling}\label{sec_assignment}

We now proceed to discuss how the PDR requirements previously derived are used to schedule the devices during a TXOP. Rather than employing a greedy method as is commonly done in wireless scheduling problems, in the proposed method we use assignment-type methods. In such assignment-type methods, we assign all scheduled devices to a PPDU and RU at the beginning of the TXOP rather than make scheduling decisions after each PPDU. To begin, we must determine a set of schedules that satisfy the constraints in \eqref{eq_c12}-\eqref{eq_c42}. Recall each device $i$ is selected to be scheduled at cycle $k$ with probability $\nu_{i,k}$ and define the set of $m_k$ devices to selected be scheduled as $\ccalI_k \subseteq \{1,2,\hdots,m\}$ where  $| \ccalI_k| = m_k$. To specify the sets of RUs that we consider in our scheduling, we first define some notation necessary in the description. We define $\hat{\ccalS}_{(n)} \subset \ccalS$ to be an arbitrary set of RUs that do not intersect over any frequency bands (i.e. satisfy the constraint in \eqref{eq_c12}) with exactly $n$ elements. To accommodate the $m_k$ devices to be scheduled, we consider a set of $S_k$ such sets  $\hat{\ccalS}_{(n_s)}$ with size $n_s$, whose combined elements total $\sum_{s=1}^{S_k} n_s = m_k$. In other words, we identify a set $S_k$ PPDUs in which the $s$th PPDU contains $n_s$ non-intersecting PPDUs. We define this full set of assignable RUs at cycle $k$ as
\begin{align}\label{eq_ru_sets}
\ccalS'_k :=\hat{\ccalS}^1_{(n_1)} \cup \hat{\ccalS}^1_{(n_2)} \cup \hdots \cup \hat{\ccalS}^{S_k}_{(n_{S_k})}.
\end{align}

Note that in \eqref{eq_ru_sets} we further superindex each set by a PPDU index $s$, in order to stress that elements are distinct between sets. That is, an RU $\bbsigma$ present in sets $\hat{\ccalS}^x_{(n_x)}$ and $\hat{\ccalS}^y_{(n_y)}$ is considered as two distinct elements in $\ccalS'_k$, denoted $\bbsigma^x$ and $\bbsigma^y$, respectively. In this way \eqref{eq_ru_sets} defines a complete set of  combinations of frequency-allocated RU and \emph{time}-allocated PPDUs to assign users during this cycle. We point out that there are numerous ways in which to define such sets of RUs in each PPDU that total $m_k$ assignments. There are various heuristic methods that may be employed to quickly identify a permissible assignment pool $\ccalS'_k$, and various simple heuristics may be developed to make this selection in a manner that reduces the overall latency of the transmission window. An example of the set $\ccalS'_k$ for scheduling $m_k = 14$ devices is shown in Table \ref{tab_rus}. 

\begin{table}[]
\centering
\begin{tabular}{|c|c|c|}
\hline 
\textbf{PPDU 1} & \textbf{PPDU 2} & \textbf{PPDU 3}        \\ \hline \hline
\multicolumn{1}{|c|}{RU 1} & \multicolumn{1}{c|}{\multirow{2}{*}{RU 10}} & \multicolumn{1}{c|}{\multirow{4}{*}{RU 13}} \\ \cline{1-1}
\multicolumn{1}{|c|}{RU 2} & \multicolumn{1}{c|}{}                       & \multicolumn{1}{c|}{}                       \\ \cline{1-2}
\multicolumn{1}{|c|}{RU 3} & \multicolumn{1}{c|}{\multirow{2}{*}{RU 11}} & \multicolumn{1}{c|}{}                       \\ \cline{1-1}
\multicolumn{1}{|c|}{RU 4} & \multicolumn{1}{c|}{}                       & \multicolumn{1}{c|}{}                       \\ \hline
\multicolumn{1}{|c|}{RU 5} & \multicolumn{1}{c|}{\multirow{4}{*}{RU 12}} & \multicolumn{1}{c|}{\multirow{4}{*}{RU 14}} \\ \cline{1-1}
\multicolumn{1}{|c|}{RU 6} & \multicolumn{1}{c|}{}                       & \multicolumn{1}{c|}{}                       \\ \cline{1-1}
\multicolumn{1}{|c|}{RU 7} & \multicolumn{1}{c|}{}                       & \multicolumn{1}{c|}{}                       \\ \cline{1-1}
\multicolumn{1}{|c|}{RU 8} & \multicolumn{1}{c|}{}                       & \multicolumn{1}{c|}{}                       \\ \hline
\multicolumn{1}{|c|}{RU 9} & \multicolumn{1}{c|}{}                       & \multicolumn{1}{c|}{}                       \\ \hline
\end{tabular}
\caption{Example of RU selection with $m_k= 14$ devices. There are a total of $S_k = 3$ PPDUs, given $n_1=9$, $n_2 = 3$, $n_3 = 2$ RUs, respectively.}
\label{tab_rus}
\end{table}

For all $i \in \ccalI_k$ and RU $\bbsigma \in \ccalS'_k$, define the largest affordable MCS given the \emph{modified} PDR requirement $\tdq_i(\hbx^{(l_i)}_{i,k})/\nu_{i,k}$ by
 \begin{align}\label{eq_mcs_select}
 \mu_{i,k}(\bbsigma) := \begin{cases}
 \max \{\mu \mid q(\bbh_{i,k},\mu,\bbsigma) \geq \tdq_i(\hbx^{(l_i)}_{i,k})/\nu_{i,k}\} \\
 1,\quad  \text{ if } q(\bbh_{i,k},\mu,\bbsigma) < \tdq_i(\hbx^{(l_i)}_{i,k})/\nu_{i,k} \ \forall \mu
 \end{cases}
 \end{align}
Observe in \eqref{eq_mcs_select} that, when no MCS achieves the desired PDR in a particular RU, this value is set to $\mu=1$ by default. The above adaptive MCS selection can be achieved based on channel conditions using the techniques outlined in~\cite{hoefel2016application}. This MCS selection subsequently then yields a corresponding time cost $\tau(\mu_{i,k}(\bbsigma), \bbsigma)$ for assigning device $i$ to RU $\bbsigma$. Further define an 3-D assignment tensor $V$---where $v^s_{ij}= 1$ when device $i$ is assigned to RU $\bbsigma^s_j$ and 0 otherwise---and $\ccalV$ as the set of all possible assignments. Recalling the form of the total transmission time given PPDU arrangements in \eqref{eq_time_slot}, the assignment that minimizes total transmission time is given by
\begin{align}\label{eq_assignment}
V^* = \argmin_{V \in \ccalV} \sum_{s=1}^S \max_{j} \left[v^s_{ij} \tau(\mu_{i,k}(\bbsigma^s_j), \bbsigma_j^s)\right].
\end{align}

The expression in \eqref{eq_assignment} can be identified as a particular form of the \emph{assignment problem}, a common combinatorial optimization problem in which the selection of mutually exclusive assignment of agents to tasks incurs some cost. Here, the cost is the total transmission time across all PPDUs necessary for scheduled devices to meet the target PDRs. Assignment problems are generally very challenging to solve---there are $m_k!$ combinations---although polynomial-time algorithms exist for simple cases. The Hungarian method \cite{kuhn1955hungarian}, for example, is a standard method for solving linear-cost assignment problems. While the cost we consider in \eqref{eq_assignment} is nonlinear, the Hungarian method may be used as an approximation. Alternatively, other heuristic assignment approaches may be designed to approximate the solution to \eqref{eq_assignment}. We note that, for the simulations performed later in this paper, we apply such a heuristic method, the details of which are left out for proprietary reasons.
{\linespread{1.3}
\begin{algorithm}[t] \begin{algorithmic}[1]
\STATE \textbf{Parameters:} Lyapunov decrease rate $\rho$
\STATE \textbf{Input:} Channel conditions $\bbH_k$ and estimated states $\hbX_{k}$
\STATE Compute target PDR $\tdq_i(\hbx^{(l_i)}_{i,k})$ for each device $i$ [cf. \eqref{eq_pdr_constraint}].
\STATE Determine selection probabilities $\nu_{i,k}$ for each device [cf. \eqref{eq_prob_c}].
\STATE Select devices $\ccalI_k$ with probs. $\{\nu_{1,k},\hdots,\nu_{m,k}\}$
\STATE Determine set of RUs/PPDUs $\ccalS'_k$ [cf. \eqref{eq_ru_sets}].
\STATE Determine maximum MCS for each device/RU assignment [cf. \eqref{eq_mcs_select}].
\STATE Schedule selected devices via assignment method.
\STATE \textbf{Return:} Scheduling variables $\{\bbSigma_k, \bbmu_k, \bbalpha_k\}$ 
\end{algorithmic}
\caption{Control-Aware Low Latency Scheduling (CALLS) at cycle $k$}\label{alg_calls} \end{algorithm}}

By combining these methods with the control-based PDR targets and selective scheduling procedure, we obtain the complete control-aware low-latency scheduling (CALLS) algorithm. The steps as performed by the centralized AP/controller are outlined in Algorithm \ref{alg_calls}. At each cycle $k$, the AP determines the scheduling parameters based on the current channel states $\bbH_k$ (obtained via pilot signals) and the current estimated control states $\hbX_k$ (obtained via \eqref{eq_state_est} for each device $i$). With the current state estimates, the AP computes target PDRs  $\tdq_i(\hbx^{(l_i)}_{i,k})$ for each device via \eqref{eq_pdr_constraint} in Step 3. In Step 4, the target PDRs are used to establish selection probabilities $\nu_{i,k}$ for each agent with \eqref{eq_prob_c}. After randomly selecting devices $\ccalI_k$ with their associated probabilities in Step 5, the set of RUs and PPDUs $\ccalS'_k$ are determined in Step 6 as in \eqref{eq_ru_sets} , based upon the number of devices selected to be scheduled $|\ccalI_k|$. In Step 7, the associated MCS values are determined each possible assignment of device to RU via \eqref{eq_mcs_select}. Finally, in Step 8 the assignment is performed using either the Hungarian method \cite{kuhn1955hungarian} or other user-designed heuristic assignment method. The resulting assignment determines the scheduling parameters $\bbSigma_k, \bbmu_k, \bbalpha_k$ for the current cycle. 
 
 \begin{remark}\label{remark_latency}\normalfont
 Observe that the CALLS method as outlined in Algorithm \ref{alg_calls} seeks to minimize the total latency of the transmission but does not explicitly prevent latency from exceeding some specific threshold $\tau_{\max}$. In practical systems, this limit may need to be enforced. In such a setting, the CALLS method can be modified so that all devices scheduled in PPDUs whose transmission end after $\tau_{\max}$ seconds do not transmit. 
 \end{remark}

\section{Simulation Results}\label{sec_simulation}
In this section, we simulate the implementation of both the control-aware CALLS method and a standard ``control-agnostic'' scheduling methods for various low-latency control systems over a simulated wireless channel. We point out the low-latency based scheduling/assignment approaches of both methods being compared are identical, with the distinguishing features being the dynamic control-aware packet delivery rates incorporated in the CALLS method. In doing so, we may analyze the performance of the control-aware design outlined in the previous section relative to a standard latency-aware approach in terms of, e.g., number of users supported with fixed latency threshold or best latency achieved with fixed number of users. As we are interested primarily in low latency settings that tightly restrict the communication resources, we consider two standard control systems whose rapidly changing state requires high sampling rates, and consequently a communication latency on the order of milliseconds. The parameters for the simulation setup are provided in Table \ref{tab_simulation}. The packet delivery rate function $q(\bbh,\mu,\bbsigma)$ is computed using the standard AWGN noise curves for wireless channels. The transmission time $\tau(\mu,\bbsigma)$ is computed in the simulations using the associated data rates of an MCS in Table \ref{tab_mcs} for a 100 byte packet and overhead (e.g. TFs) of the 802.11ax specifications. The latency overhead for this setting amounts to approximately $\tau_0 \approx 100 \mu$s.
\begin{table}
\begin{tabular}{ l | l } \hline
  Channel model & IEEE Model E (indoor) \cite{liu2014ieee} \\
  Sensor to AP distances & Random (1 to 50 meters)\\
  Transmit power & 23 dbm \\
  Channel bandwidth & 20 MHz \\
  RU sizes & 2, 4, 8, 20 MHz \\
  \# of antennas at AP & 2 \\
  \# of antennas at sensors & 1 \\
  MCS options & See Table \ref{tab_mcs} \\
  State sampling period & 10 ms \\
  \hline
\end{tabular}
\caption{Simulation setting parameters.}
\label{tab_simulation}
\end{table}

\subsection{Inverted pendulum system}

\begin{figure}
\centering
\includegraphics[height=.2\textheight]{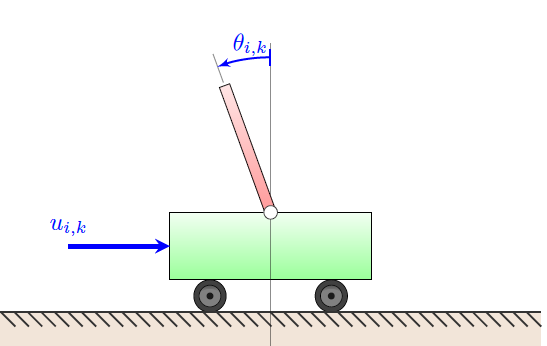}
\caption{Inverted pendulum-cart system $i$. The state $\bbx_{i,k} = [x_{i,k}, \dot{x}_{i,k}, \theta_{i,k}, \dot{\theta}_{i,k}]$ contains specifies angle $ \theta_{i,k}$ of the pendulum to the vertical, while the input $u_{i,k}$ reflects a horizontal force on the cart.}
\label{fig_inverted_pendulum}
\end{figure}

We perform an initial set of simulations on the well-studied problem of controlling a series of inverted pendulums on a horizontal cart. While conceptually simple, the highly unstable dynamics of the inverted pendulum make it a representative example of control system that requires fast control cycles, and subsequently low-latency communications when being controlled over a wireless medium. Consider a series of $m$ identical inverted pendulums, as pictured in Figure \ref{fig_inverted_pendulum}. Each pendulum of length $L$ is attached at one end to a cart that can move along a single, horizontal axis. The position of the pendulum changes by the effects of gravity and the force applied to the linear cart. For our experiments, we use the modeling of the inverted pendulum as provided by Quanser \cite{quanser}. The state is $p=4$ dimensional vector that maintains the position and velocity of the cart along the horizontal axis, and the angular position and velocity of the pendulum, i.e. $\bbx_{i,k} := [x_{i,k}, \dot{x}_{i,k}, \theta_{i,k}, \dot{\theta}_{i,k}]$. The system input $u_{i,k}$ reflects a horizontal force placed upon $i$th pendulum. By applying a zeroth order hold on the continuous dynamics with a state sampling rate of $0.01$ seconds and linearizing, we obtained the following discrete linear dynamic matrices of the pendulum system
\begin{align}\label{eq_control_orig}
\bbA_i =
\begin{bmatrix}
1 & 0 & 0 & 0 \\
0 & 2.055 & -0.722 & 4.828 \\
0 & 0.023 & 0.91 & 0.037 \\
0 & 0.677 & -0.453 & 2.055
\end{bmatrix},
\bbB_i =
\begin{bmatrix}
0.034 \\ 0.168 \\ 0.019 \\ 0.105
\end{bmatrix}.
\end{align}

Because the state $\bbx_{i,k}$ measures the angle of the $i$th pendulum at time $k$, the goal is to keep this close to zero, signifying that the pendulum remains upright. The input matrix $\bbK$ is computed to be a standard LQR-controller.

We perform a set of simulations scheduling the transmissions to control a series of inverted pendulums, varying both the latency threshold $\tau_{\max}$ and number of devices $m$. We perform the scheduling using the proposed CALLS method for control-aware low latency scheduling an, as a point of comparison, consider scheduling using a fixed ``high-reliability'' PDR of $0.99$ for all devices. Each simulation is run for a total of $1000$ seconds and is deemed ``successful'' if all pendulums remain upright for the entire run. We perform 100 such simulations for each combination of latency threshold and number of devices to determine how many devices we can support at each latency threshold using both the CALLS and fixed-PDR methods for scheduling.

\begin{figure}
\centering
\includegraphics[width=.45\textwidth]{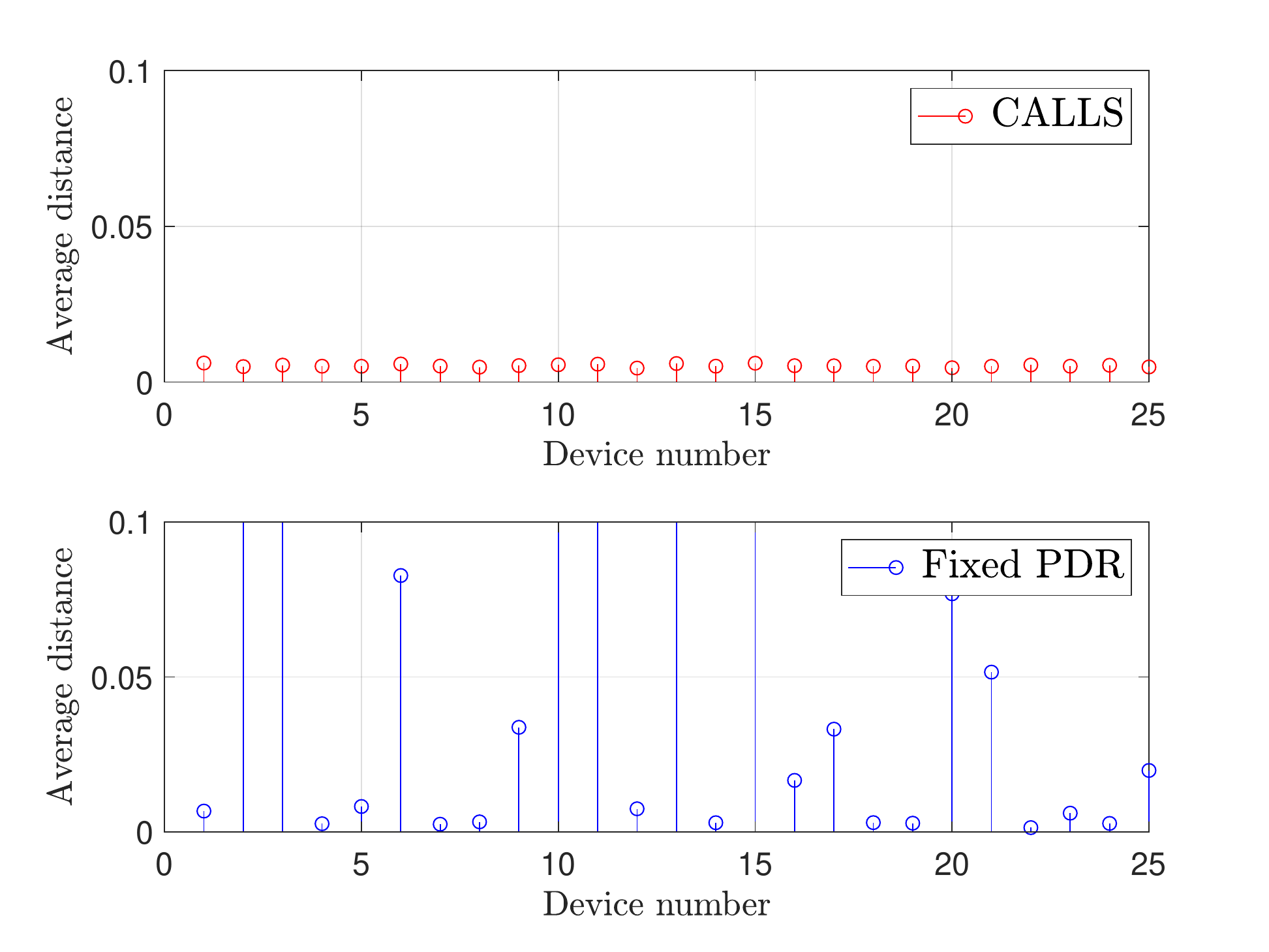}
\caption{Average pendulum distance to center vertical for $m=25$ devices using (top) CALLS and (bottom) fixed-PDR scheduling with $\tau_{\max}= 1$ ms latency threshold. The proposed control aware scheme keeps all pendulums close to the vertical, while fixed-PDR scheduling cannot.}
\label{fig_ip_dist}
\end{figure}

In Figure \ref{fig_ip_dist} we show the results of a representative simulation of the control of $m=25$ pendulum systems with a latency bound of $\tau_{\max} = 10^{-3}$ seconds. In both graphs we show the average distance from the center vertical of each pendulum over the course of 1000 seconds. In the top figure, we see by using the control-aware CALLS method we are able to keep each of the 25 pendulums close to the vertical for the whole simulation. Meanwhile, using the standard fixed PDR, we are unable to meet the scheduling limitations imposed by the latency threshold, and many of the pendulums swing are unable to be kept upright, as signified by the large deviations from the origin. This is due to the fact that certain pendulums were not scheduled when most critical, and they subsequently became unstable.

\begin{figure}
\centering
\begin{tikzpicture}[scale=0.8]
\begin{axis}[
    ybar,
    enlargelimits=0.15,
    legend style={at={(0.5,0.85)},
      anchor=north,legend columns=-1},
    ylabel={\# devices supported},
    symbolic x coords={0.5 ms,1 ms,1.5 ms},
    xtick=data,
    nodes near coords,
    nodes near coords align={vertical},
    ]
\addplot coordinates {(0.5 ms,9) (1 ms,19) (1.5 ms,36)};
\addplot coordinates {(0.5 ms,12) (1 ms,28) (1.5 ms,55)};
\legend{Fixed PDR, CALLS}
\end{axis}
\end{tikzpicture}
\caption{Total number of inverted pendulum devices that can be controlled using Fixed-PDR and CALLS scheduling for various latency thresholds.}
\label{fig_ip_bar}
\end{figure}
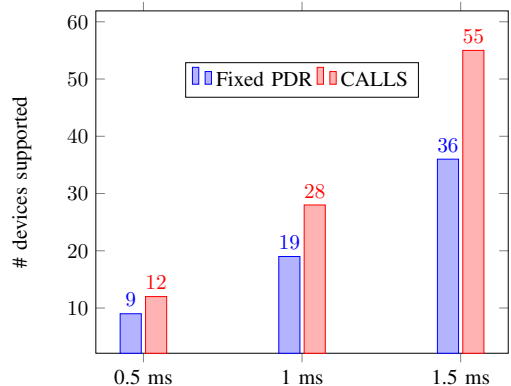

We present in Figure \ref{fig_ip_bar} the final capacity results obtained over all the simulations. We say that a scheduling method was able to successfully serve $m'$ devices if it keeps all devices within a  $|\theta_{i,k}| \leq 0.05$ error region for 100 independent simulations. Observe that the proposed approach is able to increase the number of devices supported in each case, with up to $1.5$ factor increase over the standard fixed PDR approach. Indeed, the proposed CALLS method is able to allocate the available resource in a more principled manner, which allows for the support of more devices simultaneously being controlled. 

\subsection{Balancing board ball system}

We perform another series of experiments on the wireless control of a series of balancing board ball systems developed by Acrome \cite{acrome}. In such a system, a ball is kept on a rectangular board with a single point of stability in the center of the board. Two servo motors underneath the board are used to push the board in the horizontal and vertical directions, with the objective to keep the ball close to the center of the board. The state here reflects the position and velocity in the horizontal and vertical axes, i.e. $\bbx_{i,k} := [x_{i,k}, \dot{x}_{i,k}, y_{i,k}, \dot{y}_{i,k}]$.  The input $\bbu_{i,k} = [v_x, v_y]$ reflects the voltage applied to the horizontal and vertical motors. As before, we apply a zeroth order hold on the continuous dynamics with a state sampling rate of $0.01$ seconds and linearize, thus obtaining the following dynamic system matrices,

\begin{align}\label{eq_control_orig}
\bbA_i =
\begin{bmatrix}
1 & 0.01 & 0 & 0 \\
0 & 1 & 0 & 0 \\
0 & 0 & 1 & 0.01 \\
0 & 0 & 0 & 1
\end{bmatrix},
\bbB_i =
\begin{bmatrix}
-0.0001 & 0\\ -0.02  & 0\\ 0 &-0.00008 \\ 0 & -0.01
\end{bmatrix}.
\end{align}
As before, we compute the control matrix $\bbK$ using standard LQR-control computation.

\begin{figure}
\centering
\includegraphics[width=.45\textwidth]{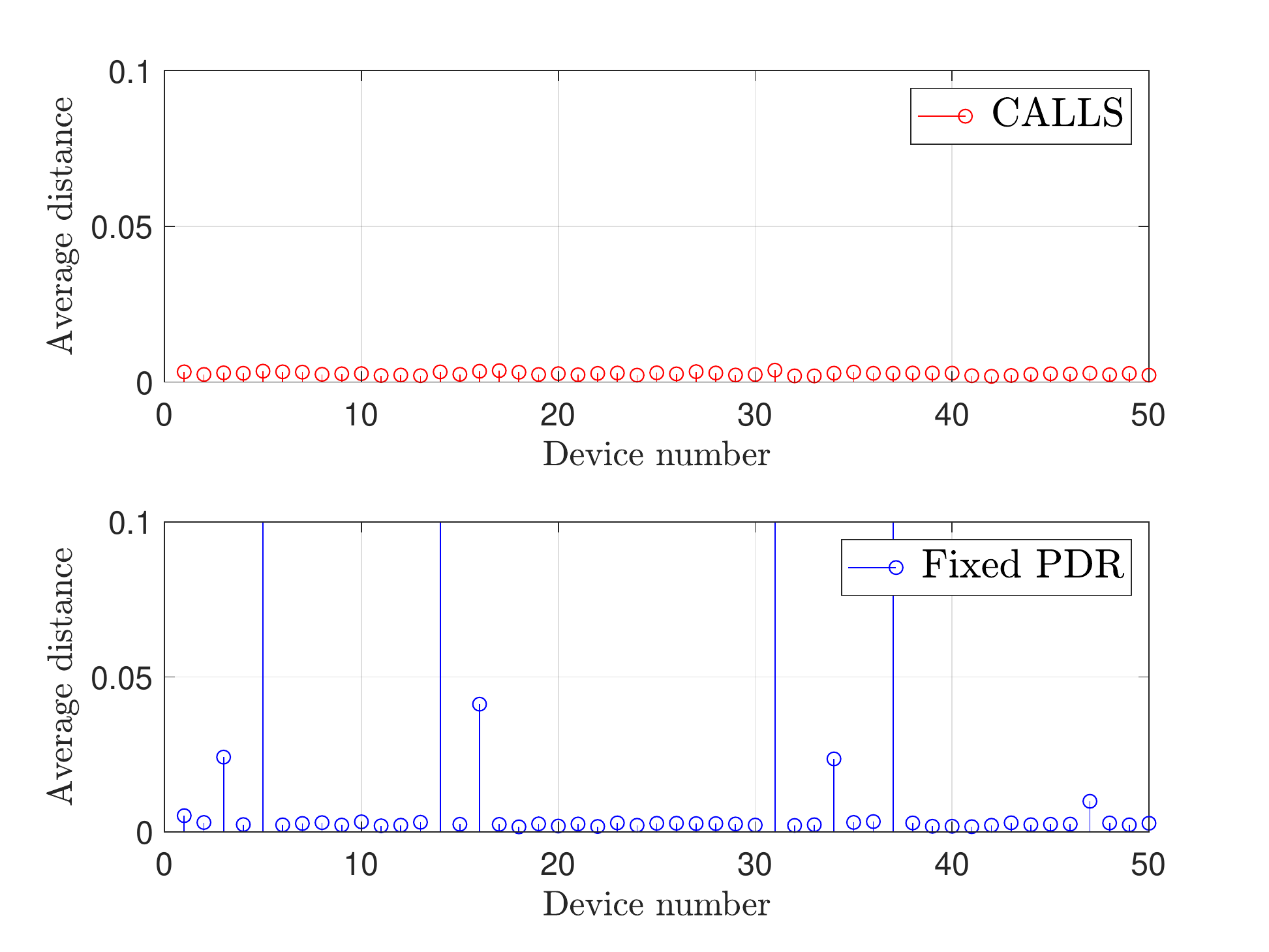}
\caption{Average ball distance to center for $m=50$ devices using (top) CALLS and (bottom) fixed-PDR scheduling with $\tau_{\max} =1$ ms latency threshold. The proposed control aware scheme keeps all balancing balls close to center, while fixed-PDR scheduling cannot.}
\label{fig_bbb_dist}
\end{figure}

In Figure \ref{fig_bbb_dist} we show the results of a representative simulation of the control of $m=50$ balancing board ball systems with a latency bound of $\tau_{\max} = 10^{-3}$ seconds. Observe that, in this system, even with a large number of users the CALLS method can keep all systems very close to the center of the board, while the fixed PDR scheduler loses a few of the balls due to the agnosticism of the scheduler.

\begin{figure}
\centering
\includegraphics[width=.45\textwidth]{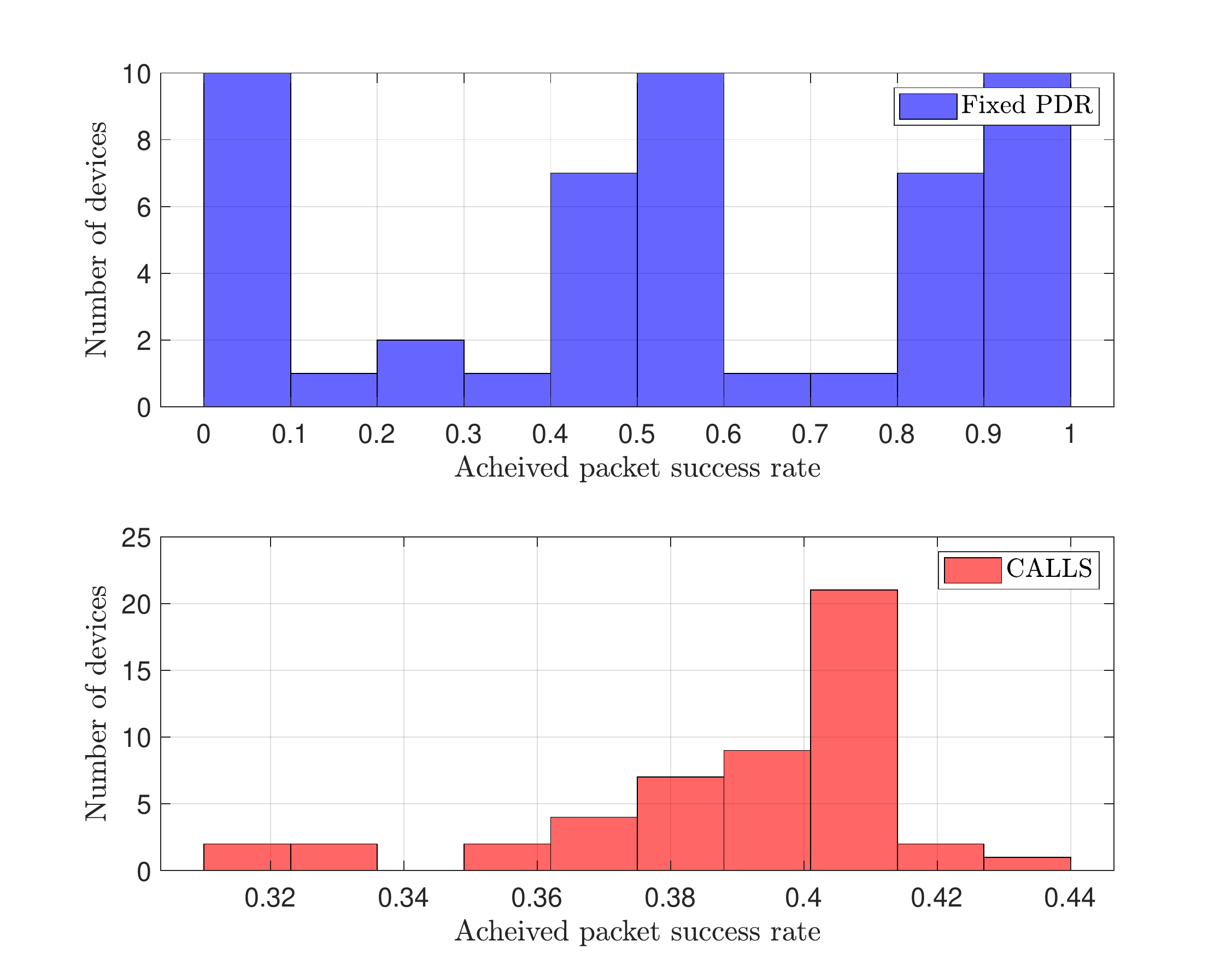}
\caption{Histogram of achieved PDRs in $m=50$ balancing board systems (top) CALLS and (bottom) fixed-PDR scheduling with $\tau_{\max} =1$ ms latency threshold. The proposed control aware scheme achieves similar PDRs for all devices, while the control agnostic scheduling results in large variation in packet delivery.}
\label{fig_bbb_psr}
\end{figure}

 To dive deeper into the benefits provided by control aware scheduling, we present in Figure \ref{fig_bbb_psr} a histogram of the actual  packet delivery rates each of the devices achieved over the representative simulation. It is interesting to observe that, in the CALLS method, the achieved PDRs are closely concentrated, ranging from 0.3 to 0.44. On the other hand, using a fixed PDR scheduling scheme, the non-variable rates are too strict for the low-latency system to support, and without control-aware scheduling the achieved PDRs range wildly from close to 0 to close to 1. In this case, some devices are able to transmit almost every cycle while others are almost never able to successfully transmit their packets. This suggests that, by using control aware scheduling, we indirectly achieve a sense of fairness across users over the long term. Further note that the PDRs required to keep the balancing board ball stable, e.g. 0.4, are relatively small. This is due to the fact that the balancing board ball features relatively slow moving dynamics, making it easier to control with less frequent transmissions. This is comparison to the inverted pendulum system, in which the pendulums were kept stable with PDRs in the range 0.6-0.75.

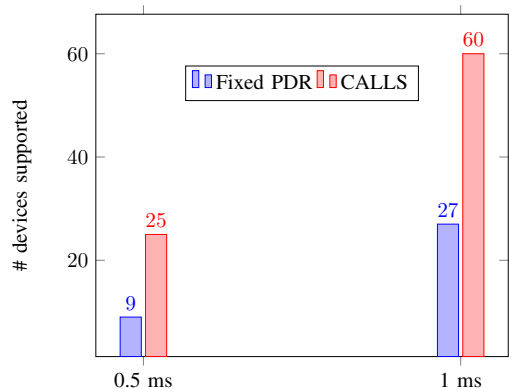
\begin{figure}
\centering
\begin{tikzpicture}[scale=0.8]
\begin{axis}[
    ybar,
    enlargelimits=0.15,
    legend style={at={(0.5,0.85)},
      anchor=north,legend columns=-1},
    ylabel={\# devices supported},
    symbolic x coords={0.5 ms,1 ms},
    xtick=data,
    nodes near coords,
    nodes near coords align={vertical},
    ]
\addplot coordinates {(0.5 ms,9) (1 ms,27)};
\addplot coordinates {(0.5 ms,25) (1 ms,60)};
\legend{Fixed PDR, CALLS}
\end{axis}
\end{tikzpicture}
\caption{Total number of balancing ball board devices that can be controlled using Fixed-PDR and CALLS scheduling for various latency thresholds.}
\label{fig_bbb_bar}
\end{figure}

We present in Figure \ref{fig_bbb_bar} the final capacity results obtained over all the simulations for the balancing board ball system. Observe that proposed approach increases the number of supported devices by factor of 2 relative to the standard fixed PDR approach. The even greater improvement here relative to the inverted pendulum simulations can be attributed to the slower dynamics of the balancing board ball, which allows for even more gains using control-aware PDRs due to the lower PDR requirements of the system.

\section{Discussion and Conclusions}\label{sec_conclusion}
In this paper we proposed a novel control-communication co-design approach to solving the radio resource
allocation problem for time-sensitive wireless control systems. Given a channel state and control state, we mathematically derive a minimum packet delivery
rate a device must meet to maintain a control-orientated target,
as defined by a stability-inducing Lyapunov function. By
dynamically assigning variable packet delivery rate targets
to each device based on its current conditions, we are able to
more easily meet feasibility requirements of a latency-constrained
wireless control problem and maintain stability
and strong performance. We perform simulations on numerous
well-studied low-latency control problems to demonstrate the
benefits of using the control-aware approach, which can include
a 2x gain on number of devices that can be supported.

The results presented in this paper suggest an interesting
potential for control-aware resource allocation and scheduling, particularly in low-latency industrial systems. By considering the control-specific targets such as maintaining
stability or an error margin, we observe that the standard
high reliability targets considered (e.g. packet delivery rates $\geq 0.999$) can in some cases be substantially stricter than
necessary for adequate performance. Wireless control systems
with sufficiently slow dynamics can be kept stable with much
lower packet delivery rates, which in turn make low-latency
communications more achievable. Furthermore, in realistic industrial systems there will be many heterogeneous devices being controlled, whose variation in communication needs is well-served by control-aware opportunism proposed in this paper. This suggests the potential for wireless communications to be adopted using a smart control-communication co-design approach even
while ultra-reliable wireless system technology remains under development.

\urlstyle{same}
\bibliographystyle{IEEEtran}
\bibliography{wireless_ll_control,scheduling_control}

\end{document}